\documentclass{elsart3p}

\usepackage{graphicx}

\usepackage{amssymb}

\begin{document}

\begin{frontmatter}



\title{Integrated Sachs-Wolfe effect in the era of precision cosmology}


\author{Levon Pogosian\corauthref{cor1}}

\address{Department of Physics, Syracuse University, Syracuse, NY 13244-1130, USA}

\corauth[cor1]{From September 1, 2006, Department of Physics, Simon Fraser University, Burnaby, BC  V5A 1S6, Canada}

\begin{abstract}
Recent detections of the Integrated Sachs-Wolfe effect through
the correlation of the cosmic microwave background temperature anisotropy with
traces of large scale structure provided independent evidence for
the expansion of the universe being dominated by something other than
matter. Even with perfect data, statistical errors will limit the accuracy of 
such measurements to worse than $10$\%.
On the other hand, the extraordinary sensitivity of the ISW effect to
the details of structure formation should help to make up
for the lack of precision. In these conference proceedings I discuss the extent 
to which future ISW measurements can help in testing the physics 
responsible for the observed cosmic acceleration.
\end{abstract}

\end{frontmatter}


As the cosmic microwave background (CMB) photons travel to us from the surface of
last scattering they pass through gravitational potentials created by accreting matter. 
Photons blueshift when they fall into potential wells and redshift as they climb out. 
In a matter dominated Freidman-Robertson-Walker (FRW) universe the potentials 
remain constant with respect to the
co-expanding coordinates. Hence, during matter domination photons redshift
with the expansion but do not gain or loose additional energy after passing through 
the potentials. However, any deviation from matter domination, e.g. due to 
dark energy or curvature, causes the potentials to evolve with time, 
leading to a net change in photon energies as they
pass through them. This is seen as an additional
CMB temperature anisotropy called the Integrated Sachs-Wolfe (ISW) effect \cite{SW67}.

The ISW contribution to the CMB temperature anisotropy in a direction ${\hat n}$ on the
sky is approximately given by
\begin{equation}
\Delta^{ISW}({\hat n}) \approx -2\int d\eta \  {\dot \Phi}[r\hat{n},\eta] \ ,
\label{isw_def}
\end{equation}
where $\Phi$ is the Newtonian potential, the dot denotes a derivative with respect
to the conformal time $\eta$, and $r(\eta)$ is the proper distance. 
This expression ignores the effects of reionization and the possibility of
a non-zero anisotropic stress, both of which are negligible in conventional 
cosmological scenarios\footnote{The anisotropic stress can play an
important role in some of the alternative models of gravity \cite{zhang06}}. 
The integral in (\ref{isw_def}) ranges from the time of last
scattering until today. However, it may be possible to isolate the ISW generated 
over a smaller period of time by, e.g., correlating CMB with large scale
structure (LSS). Furthermore, one can imagine having a measurement 
of the ISW effect in 
multiple redshift bins. This would provide a probe of how ${\dot \Phi}$ evolves with
time. It was shown in \cite{CHB03} that knowing ${\dot \Phi}$ in the $0<z<2$ range
with $10$\% accuracy in redshift bins of width $0.1$ would
provide constraints on dark energy parameters comparable to those from $1$\%
accuracy measurements of other functions of redshift, such as the luminosity distance
or the growth factor.

\begin{figure}
\centering
\includegraphics[width=80mm]{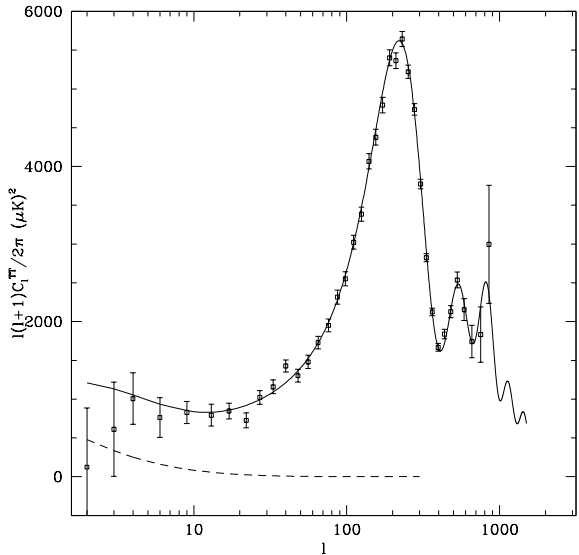}
\caption{\label{isw_cl} The total CMB angular power spectrum (solid line) and
the ISW contribution (dashed lined) for the $\Lambda$CDM model plotted on top of
the WMAP 1-year data. The error bars at low $\ell$ are dominated by cosmic variance,
which makes it difficult to isolate the ISW contribution to the spectrum.}
\end{figure}

It is practically impossible to measure the ISW effect using CMB measurements alone. 
As shown in Fig.~\ref{isw_cl} for the case of the standard $\Lambda$CDM
model, the ISW contribution to the total CMB spectrum is only significant on the largest scales,
where our ability to extract information is fundamentally limited by cosmic variance.
Fortunately, there are ways to detect the ISW effect by means other than 
the CMB spectrum. One possibility, suggested in \cite{KL97}, is to look at the CMB polarization 
towards clusters of galaxies. Polarization is produced when photons with a
non-zero temperature quadrupole scatter on charged particles,
such as hot gas in clusters. Hence, measuring the
polarization of CMB coming from the direction of a cluster is telling us 
what the CMB temperature quadrupole was at the redshift of the cluster.
The main contribution to the temperature quadrupole comes from the epoch of
last scattering and does not change with redshift. The ISW contribution,
on the other hand, will evolve with time. Therefore, knowing the quadrupole at several
redshifts can help isolate its time-varying component. 
Estimates of how well the future generation of experiments, such as CMBPOL, 
coupled with cluster redshift surveys, can isolate the ISW component using this technique
were presented in \cite{CHB03} along with preliminary forecasts on dark energy parameters.
The potential and the limitation of this type of measurements have been further studied in
\cite{portsmouth04,SP05,Bunn06}.

Another way to isolate the ISW effect from the rest of the CMB is to cross-correlate the 
large scale CMB anisotropy with a map of the CMB shear field. This idea, proposed in \cite{KKC03}, 
is based on the fact that the same gravitational 
potentials that lens the CMB would also produce an ISW signal. So, there will be a non-zero 
correlation of shear with the ISW part of CMB, but not with the primordial part coming 
from the surface of last scattering. As shown in \cite{KKC03}, if the low CMB large scale power 
was due to a cutoff in the primordial spectrum, the signal to noise of this correlation 
would be significantly enhanced. This is because the correlation of 
the ISW part of CMB with the shear would not be affected, but the reduction in the
primordial part of the CMB spectrum would lower the variance. The forecast, according
to \cite{KKC03}, is that this type of measurements will be possible with future missions such
as CMBPOL, but probably not before then.

At this time, the most feasible way of measuring the ISW effect is by 
correlating CMB with large scale structure, as first proposed by Crittenden and 
Turok in 1995 \cite{CT96}. Such correlation has indeed been detected
between the CMB data from WMAP and existing catalogs of tracers 
of large scale structure \cite{detection}. The combined significance of
the detection of the ISW effect using current data from SDSS, 2MASS and NVSS 
is, quite conservatively, at a $5\sigma$ level \cite{Ryan}.

Current measurements of the CMB/LSS correlation are only beginning to approach 
precision levels where they can contribute information about cosmological parameters.
For example, in Ref. \cite{CGM05} the compiled cross-correlation data from \cite{GMM04} 
was used to put constraints on a constant dark energy equation of state parameter 
$w$ and the dark energy speed of sound $c_s^2$. The verdict is that
these constraints are marginally informative but far from competitive.
The same complication of data was used in \cite{CCGM05} to put a
constraint on the tensor mode amplitude $r$. Since LSS only correlates with 
the scalar modes, does not depend on the reionization optical depth,
and provides a handle on the galaxy bias, it helps to break some degeneracies
that affect the extraction of $r$. It was found in \cite{CCGM05} that adding the
existing cross-correlation data to the information pool lowers the bound $r<0.9$
derived from the WMAP 1-year data combined with SDSS or 2dF to $r<0.5$.
However, as the quality of the data improves, the relative utility of the 
cross-correlation will become weaker. Already, it cannot significantly improve 
on the constraint $r<0.3$ obtained from the WMAP 3-year data combined with SDSS \cite{wmap3}.

\begin{figure}
\centering
\includegraphics[width=80mm]{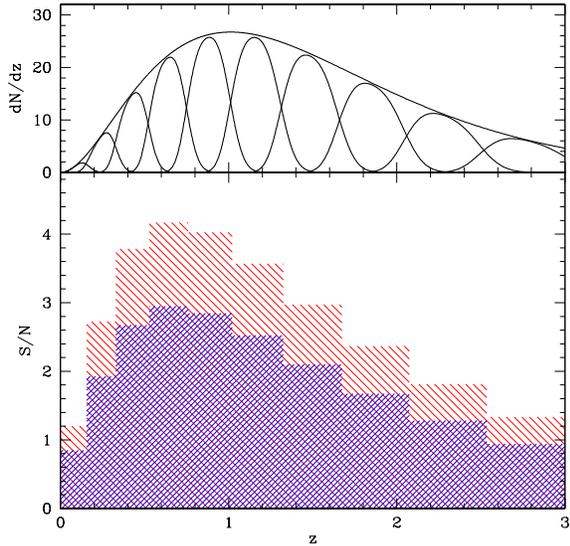}
\caption{\label{lsst_bins} {\bf Top panel:} the total expected
LSST galaxy number distribution and 
its breakdown into ten photometric redshift bins. 
The distribution shown here is plotted using
the parameters provided in the LSST white paper \cite{lsst}, and is
slightly different from the one assumed in \cite{CCGM05}. This difference does not 
lead to a noticeable change in any of the constraints on dark energy parameters. 
{\bf Bottom panel:} the expected LSST/Planck cross-correlation signal-to-noise in 
each of the bins for the best fit LCDM model. The dark (red) shading corresponds
to a half sky coverage by LSST, while the light (red) shading is for the full sky.}
\end{figure}
What we would like to address in the remainder of these proceedings is the utility of 
future measurements of the ISW effect. We will imagine correlating CMB data from Planck
with photometric redshift galaxy catalogs from the proposed Large Synoptic 
Survey Telescope (LSST) \cite{lsst}. The details of our assumptions 
for Planck and LSST can be found in \cite{PCSCN05}. We assume that LSST will cover
half of the full sky and catalog around $50$ gal/arcmin$^2$ in ten photometric
redshift bins spaced evenly between $z=0$ and $3$. The total galaxy number distribution
and its breakdown into photometric bins are plotted as a function of redshift in
the top panel of Fig.~\ref{lsst_bins}. Correlating CMB with galaxies in each of the photometric 
bins gives the ISW contribution at mean redshifts of different bins, allowing us
to map the evolution of $\dot{\Phi}$ with time.

In order to put the discussion on a more quantitative footing we need to define some
notation \cite{GPT04}. The CMB temperature is a function of the direction on the sky, $T(\hat{n})$.
Its two-point auto-correlation function $C(|\hat{n}-\hat{n}'|)=C(\theta)=
\langle T(\hat{n})T(\hat{n}')\rangle$
is usually decomposed into a Legendre series with coefficients $C_\ell$. It is
common to evaluate the CMB angular spectrum, $\ell(\ell+1)C_\ell/2\pi$ vs $\ell$, which
is the quantity plotted in Fig.~\ref{isw_cl}. The distribution of galaxies in
the $i$-th photometric bin is also a function of $\hat{n}$, $M^{(i)}(\hat{n})$. Here
one can define Legendre coefficients $M^{(ij)}_\ell$ corresponding to the two-point 
correlation between galaxies in the $i$-th and the $j$-th bins. The cross-correlation
between CMB and galaxies in one of the bins can be similarly represented in terms of
the angular spectrum $X^{(i)}_\ell$ defined via
\begin{equation}
\langle T(\hat{n}) M^{(i)}(\hat{n}') \rangle  \equiv X^{(i)}(\theta)
= \sum_{\ell=0}^{\infty} {2\ell+1 \over 4\pi} X^{(i)}_\ell P_\ell(\theta)
\end{equation}
\begin{figure}
\centering
\includegraphics[width=80mm]{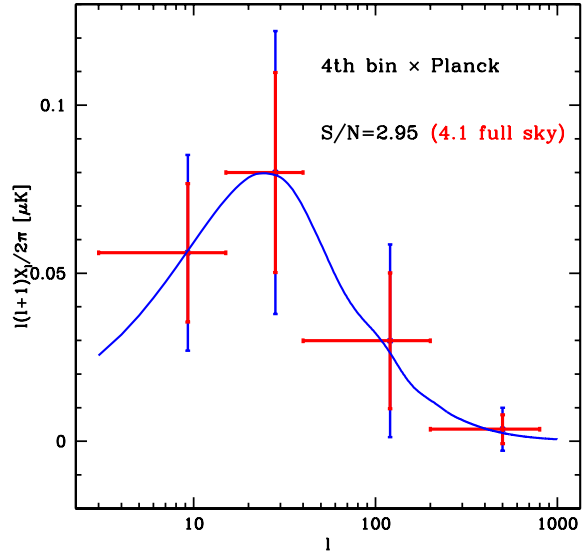}
\caption{\label{xl} The angular spectrum of the correlation of the Planck CMB temperature with
the $4$-th LSST bin in Fig.~\ref{lsst_bins} for the best fit LCDM model. Shown are the
statistical error bars based on a half (lager blue) and a full (smaller red)
sky coverage by LSST.}
\end{figure}
The time-evolution of galaxy clustering is described by 
the so-called growth factor, $D(a)$, which in general relativity
is scale-independent in the linear regime. The ISW contribution
to the CMB anisotropy is sourced by the time-derivative of the gravitational
potential, related to the growth factor via the Poisson equation: 
$\dot{\Phi}\propto d(D(a)/a)/d\eta$. The correlation between clustering
and the CMB is essentially probing the quantity 
\begin{equation}
P(\bar{z}_i)=D(\bar{z}_i)\left({d \over dz}[(1+z)D(z)]\right)_{z=\bar{z}_i} \,
\end{equation}
where $\bar{z}_i$ is the mean redshift of a given bin. Having multiple bins,
as in Fig.~\ref{lsst_bins}, would allow one to trace the evolution of $P(z)$
over a wide range of redshifts. Since, as shown in \cite{CHB03,PCSCN05,Pogosian05}, it is 
an extremely sensitive probe of dark energy, even a marginal measurement of
$P(z)$ would be of high value.

The statistical signal-to-noise (S/N) in $X^{(i)}_\ell$ is given by
\begin{equation}
\left({S \over N}\right)^2_\ell = {f_{\rm sky} (2\ell+1) \over 4\pi }
{(X^{(i)}_\ell)^2 \over C_\ell M_\ell^{(ii)}+ (X^{(i)}_\ell)^2} \ ,
\end{equation}
and generally $C_\ell M_\ell^{(ii)} \gg (X^{(i)}_\ell)^2$ because there is
a large contribution to the $C_\ell$ coming from the last scattering epoch
that does not correlated with the low-redshift universe. The S/N
expected for the LCDM model in each of the LSST bins is shown in the bottom
panel of Fig.~\ref{lsst_bins}.
\begin{figure}
\centering
\includegraphics[width=80mm]{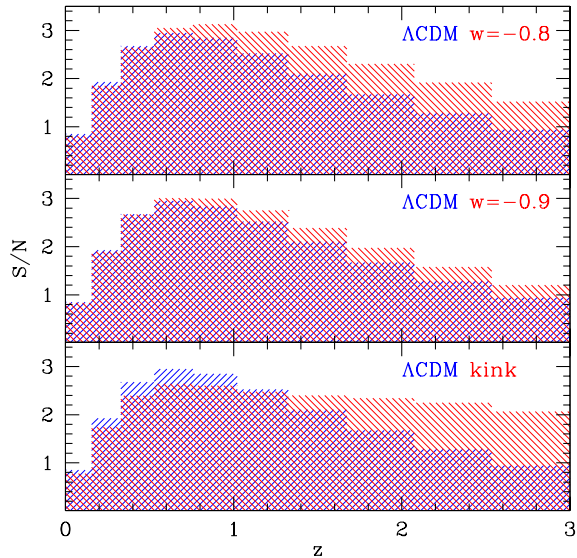}
\caption{\label{sn_3} The expected signal-to-noise in correlation of the Planck CMB 
temperature with the LSST galaxies (Fig.~\ref{lsst_bins})
for three different dark energy models ($135^\circ$ red shading) compared to 
the best fit LCDM ($45^\circ$ blue shading). A half-sky LSST coverage is assumed.
Shown are results the constant $w=-0.8$ and 
$w=-0.9$ models, and the kink model (Fig.~\ref{eos_kink}). The Hubble parameter was adjusted
to give nearly identical CMB spectra for all models being compared.}
\end{figure}
The distribution of the S/N can depend strongly on the particular
dark energy model. In models with $w > -1$ dark energy starts to dominate
earlier, leading to an earlier contribution to the ISW effect. The difference in the S/N
for different models is shown
in Fig.~\ref{sn_3}, where the Hubble parameter in each of the three case was adjusted 
to keep the peak structure of the CMB spectra nearly identical to the best fit LCDM.
The difference is particularly pronounced for the
so-called kink model, in which $w$ undergoes a sharp transition from a higher
to lower value.

\begin{figure}
\centering
\includegraphics[width=85mm]{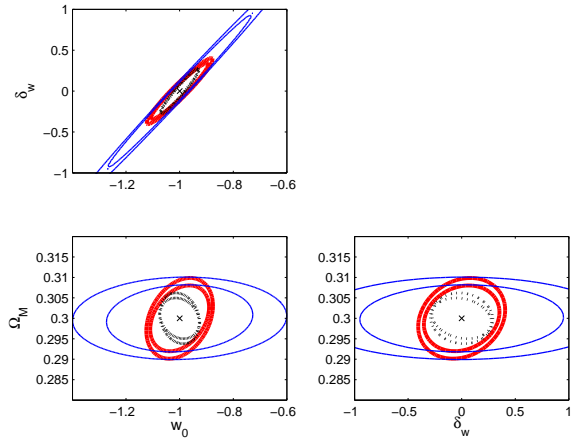}
\caption{\label{linear} The $1\sigma$ projected Fisher contours for
parameters of the model with $w$ given by eq.~(\ref{m1}). Other
parameters were marginalized over. The (blue) thin solid lines
are for Planck's CMB. Here, the larger contours are for temperature
spectra alone, while the smaller ones include information from
CMB polarization. The (red) think solid lines are constraints from
CMB spectra combined with CMB correlation with LSST galaxies. The
(black) dashed contours are for the supernovae from SNAP (using
the assumptions of \cite{linder_snap}) combined with
Planck. The plot is taken from \cite{PCSCN05}.}
\end{figure}
In \cite{PCSCN05} Fisher matrix analysis was used to forecast
errors on evolving $w(z)$. 
Such forecasts, of course, depend on the choice of the model of $w(z)$ 
and the fiducial values of the parameters.
Here we show results for two models. In Model I the parameterization is \cite{wa}
\begin{equation}
w(a)=w_0+(a-1)\delta_w
\label{m1}
\end{equation}
with fiducial values of the parameters chosen to represent the LCDM model:
$w_0=-1$, $\delta_w=0$. The projected $1\sigma$ contours on these parameters and
$\Omega_M$ are shown in Fig.~\ref{linear}. Spatial flatness and adiabatic initial
conditions, favored by the WMAP data, were assumed.
The usual cosmological parameters
were also varied and marginalized over. It is clear that for this
model the cross-correlations constraints on dark energy are informative, but 
somewhat weaker than those derived from the supernovae data. 

\begin{figure}
\centering
\includegraphics[width=70mm]{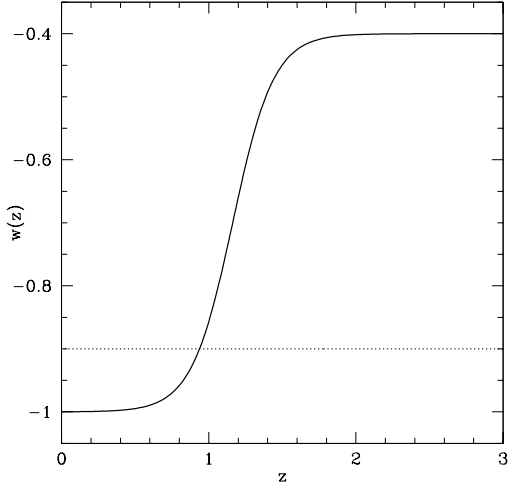}
\caption{\label{eos_kink} The $w(z)$ vs $z$ in the kink model with $\Delta_z=0.3$,
$\Delta_w=-0.3$ and $\langle w \rangle=-0.9$ This model predicts CMB spectrum
nearly identical to that with $w=const=\langle w \rangle$, shown with a dash line.}
\end{figure}

To illustrate how the choice of the model can affect the forecasts, we
also show results obtained in \cite{PCSCN05} for the so-called kink model.
In the kink model, shown in Fig.~\ref{eos_kink}, $w$ transitions from $-1$
at low redshifts to a higher value in the past. The supernovae would have
a difficulty "seeing" such a transition had it occurred at a sufficiently
high redshift. The CMB/LSS correlation, on the other hand, can prove to be 
a useful tool in this case. We describe the kink in Fig.~\ref{eos_kink} using
three parameters. Those are the width
of the transition $\Delta_z $, the total change in the equation of state,
$\Delta_w$, and the weighted average value $\langle w \rangle$ defined as
\begin{equation}
\langle w \rangle ={\int da \Omega_{DE}(a) w(a) \over \int da \Omega_{DE}(a)}
\end{equation}
This quantity controls the angular diameter distance to the last scattering \cite{wcmb}.
Making $\langle w \rangle$ and explicit parameter has the advantage
of being able to clearly separate models that fit the peak structure of
the CMB angular spectrum from those that do not.

\begin{figure}
\centering
\includegraphics[width=85mm]{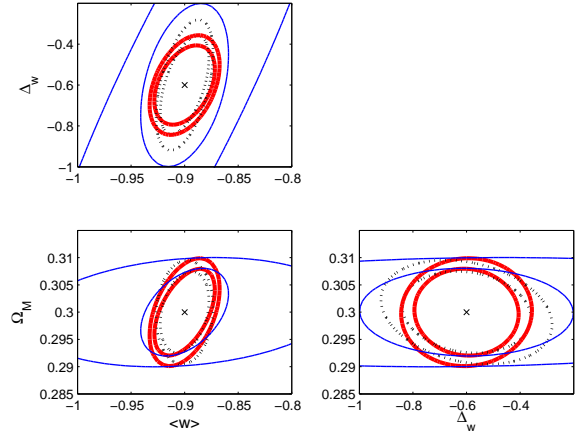}
\caption{\label{kink} The $1\sigma$ projected Fisher contours for parameters
of the kink model. $\Delta_z$ and other parameters were marginalized over. The
meaning of the different lines is the same as in Fig.~\ref{linear} -- the
contours corresponding to the cross-correlation of Planck and LSST, with added
information from CMB spectra, are shown with (red) thick solid lines.}
\end{figure}

Fisher errors for the kink model parameters were calculated in \cite{PCSCN05}
under the assumption of flat geometry and adiabatic initial conditions. In
Fig.~\ref{kink} we show the $1\sigma$ projected contours on $\Omega_M$, $\Delta_w$
and $\langle w \rangle$. The other parameters, including $\Delta_z$, are
marginalized over. For this model the cross-correlation measurements
can provide competitive constraints on dark energy evolution.

A more model-independent approach to forecasting constraints on $w(z)$ is
provided by the principal component analysis introduced to dark energy
studies in \cite{HS02}. A study comparing the power of different types
of dark energy experiments using the principal component approach was performed
in \cite{CP05}. It was shown that the CMB/LSS correlation can provide a 
constraint on one principal component of $w(z)$ which would be complementary to 
information provided by other probes. The CMB/LSS correlation, similar to the CMB
auto-correlation, is sensitive to an averaged value of $w(z)$ with added weight at
higher redshifts.

Constraining the dark energy equation of state is only one of several 
promising applications of the ISW effect. It can also
be a potentially useful probe of dark energy clustering \cite{BD04,HS04,CGM05}, and
a unique tool for testing models that 
explain apparent cosmic acceleration through modifications of Einstein's 
theory of gravity \cite{zhang06,SSH06}.

In summary, the ISW effect is not the most precise measurement we have in
modern cosmology. However, with future surveys of large scale structure, 
such as LSST and the Square Kilometer Array (SKA) \cite{ska}
covering large fractions of the sky and a wide range of redshifts, cross-correlation
with CMB will likely become a standard tool for testing cosmological models
that predict unconventional growth history at high to moderate redshifts.
What allows the ISW effect
to overcome the lack of accuracy is its extraordinary sensitivity to
the details of the process of structure formation, whether it involves a scalar
field dark energy, dark energy clustering or modifications of gravity. The ISW effect 
is free of some of the complicated physics involved in other probes of structure 
growth as it probes only linear scales, has a linear dependence on the galaxy bias,
probes deep in redshift and is free of some of the degeneracies that hinder 
CMB studies on large scales. Given these strengths, 
and the fact that the utility of the
ISW effect is a relatively under-researched subject, its most useful application
may still be discovered in the future.



\

\centerline{\bf Acknowledgments}
This talk was based on collaborations with 
R.~Crittenden, P-S.~Corasaniti, J.~Garriga, R.~Nichol, C.~Stephan-Otto and
T.~Vachaspati. I would like to thank A.~Cooray, M.~Kaplinghat and all others
involved in organizing the Fundamental Physics with CMB workshop at Irvine
for putting togheter this timely and enjoyable meeting.

\end{document}